\documentclass[11pt,twoside]{article}
\usepackage{asp2010}

\resetcounters

\begin{document}

\title{Linking Publications and Observations - the ESO Telescope Bibliography }
\author{Silvia Meakins and Uta Grothkopf$^1$
\affil{$^1$European Southern Observatory (ESO), Karl-Schwarzschild-Str. 2, 85748 Garching near Munich, Germany}}

\begin{abstract}
Bibliometric studies have become increasingly important in evaluating individual scientists, specific facilities, and entire observatories. In this context, the ESO Library has developed and maintains two tools: FUSE, a full-text search tool, and the Telescope Bibliography (telbib), a content management system that is used to classify and annotate ESO-related scientific papers. 

The new public telbib interface provides faceted searches and filtering, autosuggest support for author, bibcode and program ID searches, hit highlighting as well as recommendations for other papers of possible interest. It is available at\\ \url{http://telbib.eso.org}.
\end{abstract}

\section{Introduction}
Telescope bibliographies are important tools to measure scientific output. Typically, they contain all (or all refereed) papers using observational data from specific facilities that were published in the scholarly literature. 

Telescope bibliography databases are therefore the ideal source to derive various kinds of reports and statistics. For instance, management and governing bodies of observatories may be interested in publication and citation statistics to evaluate the performance of observatories and telescopes. Instrument scientists often need reports regarding the scientific impact of specific instruments and research programs. Such reports can also provide guidelines for future telescopes and instruments. For the astronomy community at large, it is important that telescope bibliographies interconnect resources: publications are linked to the observing programs that generated the data and to the actual data in the archive, and in turn scientists will be able to go from the archival data directly to all publications that use these data.  

In this context, the ESO Library has developed and maintains two tools: 
(1) FUSE is a full-text search tool that semi-automatically scans defined sets of journal articles for organizational keyword sets, while providing highlighted results in context;
(2) the Telescope Bibliography (telbib) is used to classify ESO-related papers, store additional metadata, and generate statistics and reports. Both tools rely heavily on the NASA ADS Abstract Service for bibliographic metadata.

In this paper, we describe how FUSE and telbib link publications and observations and explain the main features of the new public telbib interface.

\section{FUSE and telbib at ESO}
FUSE and telbib form part of a workflow that links published literature with data located in the ESO archive. The result is an information system that answers predominantly two questions: 
\vspace{-.2cm}

\begin{itemize}
\item Which ESO facilities generated the data used in the scientific literature?
\vspace{-.3cm}

\item Which publications used data provided by specific ESO facilities?
\end{itemize}

\vspace{-.2cm}

\noindent
In the following, essential components of FUSE and telbib are explained (see Fig. 1). 

\subsection{Access to literature}
In order for FUSE to work, it is necessary to have access to the electronic versions of all scientific journals that shall be monitored. FUSE provides search methods for these journals that allow to retrieve the full-texts of articles (typically in PDF format) from the publishers' websites.

\subsection{FUSE search tool}
FUSE is a PHP/MySQL tool created by the ESO Library. It converts PDFs into text files and scans them for user-defined keyword sets. If any keywords are detected in the text files, they are highlighted and shown in context on the results page. 

\subsection{Telbib back-end}
After FUSE identifies possible candidates for the ESO Telescope Bibliography, these papers are inspected visually in detail. Records that shall be added to telbib are imported into the database through the librarians' interface (telbib back-end, implemented with PHP/Sybase). Bibliographic information (authors, title, publication details) along with further information like current number of citations, author-assigned keywords, and author affiliations are imported from the ADS Abstract Service. 

\subsection{Telbib tags and program IDs}
The ESO librarians extensively tag and annotate each telbib record. Such tags include standardized descriptions of telescopes, instruments, surveys, and other information. Most importantly, all ESO program IDs that provided data are assigned to the telbib record of the paper. 

\subsection{Data archive}
The program IDs assigned to telbib records provide links to the corresponding data in the ESO archive. In this way, readers of scientific papers who are interested in the data used in the publication can easily find the observing programs that were used in the research. The data can be requested after the usual one year proprietary period. 

\subsection{Telbib front-end} 
Program IDs in the ESO archive provide links back to telbib, listing all scientific papers that use specific observing programs. Of course telbib can also be queried directly through the public user interface. More information on the telbib front-end is given in the following section. 

\begin{figure}
\plotone{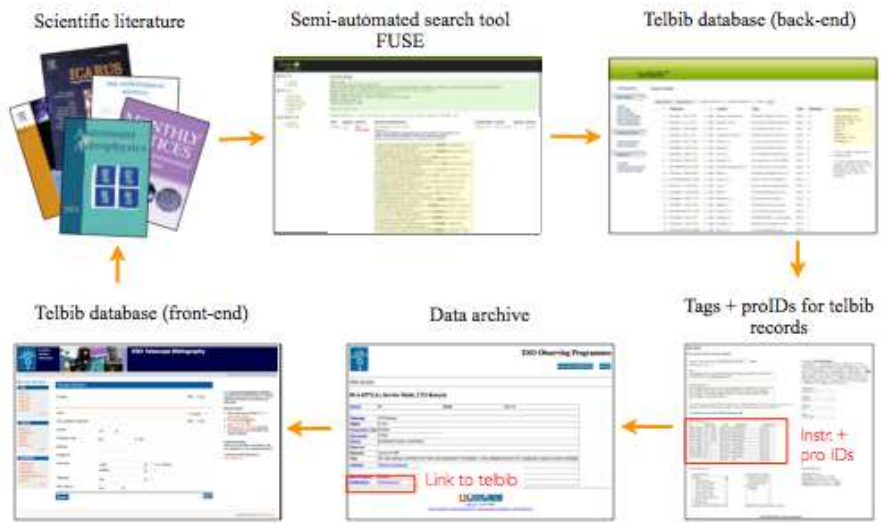}
\caption{FUSE and telbib link published literature and data located in the archive, keeping track of the observing facilities and programs used for the research.}
\end{figure}

\section{The new public telbib interface}
For many years, the public interface of the ESO Telescope Bibliography had not changed its look and feel. Now, telbib's front-end has undergone a complete makeover. It has been created and will be further developed by the ESO librarians. 

A state-of-the-art interface has been implemented. Using Apache Solr together with PHP, telbib now provides new features and sophisticated search functionalities.
The new system will be rolled out in the final quarter of 2011. It will be accessible at \url{www.eso.org/libraries/telbib.html}. 

\subsection{Search interface}
The telbib front-end interface provides a variety of options to query the database. These include searches by bibliographic information (authors, title words, author-assigned keywords, publication year, etc.) as well as by observing facilities (instruments, telescopes) and program IDs. The main search screen shows a list of the top 5 journals and instruments, indicating the number of records in telbib for each journal and using data from these instruments, respectively. In addition, the most recent five years are displayed along with the number of papers per year. 

A spellchecker is available for certain search fields (for instance author names, title words). In case search terms entered in these fields do not lead to any hit, the system will provide hints towards search terms that will turn up results ("Did you mean...?"). In addition, queries for authors, bibcode, and program IDs are supported by an autosuggest feature that offers search terms which exist in the index after at least two characters have been entered. 

\subsection{Results page}
The results page lists papers that fulfill the query parameters in six columns, showing the publication year, first author, instruments, program IDs, and the bibcode of each paper. Titles are linked to the detailed view of records, program IDs lead to the ESO observing schedule and ultimately to the archive from where the data can be requested, and bibcodes are connected with the full-texts at ADS. 

In order to limit search results, faceted filtering is available in the "Refine Search" area on the left-hand side. Facets exist for publication years, journals, and instruments. For the latter two filters, the top 5 among these results are shown, together with the five most recent years. The lists can be expanded by clicking on the "More..." button. 

In order to further use results sets, they can be exported into comma-separated (.csv) or tab-separated (.txt) format.  

\begin{figure}
\begin{center}
\includegraphics[width=0.65\textwidth]{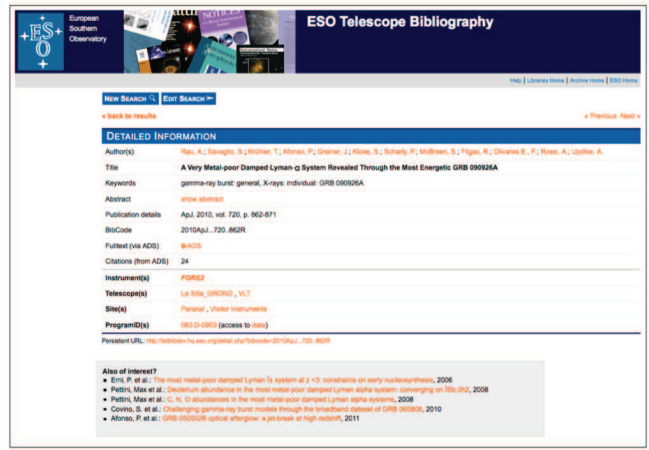}
\caption{The new public interface of ESO's telbib: detailed record display}
\end{center}
\end{figure}

\subsection{Detailed record display}
The detailed record view (Fig. 2) shows all ESO observing facilities that were used in the paper as well as additional tags like survey names. Search terms are highlighted for easy identification. Instruments, telescopes, and observing sites are hyperlinked and will retrieve other papers that use the same facilities. 

For program IDs, two links are offered. Clicking on the program ID itself will evoke a new search for all papers that use the given program. Selecting "access to data" takes users to the observing schedule and from there to the archive where the respective data can be requested if the proprietary period has ended. 

At the bottom of the page, users will find recommendations for other papers that may be of interest. This section is entitled "Also of interest?" and offers access to other papers with similar content than the one currently displayed .

\acknowledgements We are very grateful to Chris Erdmann, now at the Harvard-Smithsonian CfA, who created the first versions of FUSE and telbib. Both programs make extensive use of NASA's Astrophysics Data System. Many thanks to all of them.

\end{document}